\documentclass[aps,prb,twocolumn,superscriptaddress]{revtex4-1}
\usepackage{graphicx}
\usepackage{array}
\usepackage{dcolumn}
\usepackage{bm}
\usepackage{float}
\usepackage{enumitem}

\newcommand{\FGT}{Fe$_{5}$GeTe$_2$~}
\newcommand{\FGTx}{Fe$_{5-x}$GeTe$_2$~}

\newcommand{\TC}{$T_{\rm C}$~}

\newcommand{\MT}{$M$($T$)~}
\newcommand{\MH}{$M$($H$)~}

\begin{document}


\title{Physical properties and thermal stability of Fe$_{5-x}$GeTe$_2$ single crystals}

\author{Andrew F. May}
\email{mayaf@ornl.gov}
\affiliation{Materials Science and Technology Division, Oak Ridge National Laboratory, Oak Ridge, TN 37831}

\author{Craig A. Bridges}
\affiliation{Chemical Sciences Division, Oak Ridge National Laboratory, Oak Ridge, TN 37831}

\author{Michael A. McGuire}
\affiliation{Materials Science and Technology Division, Oak Ridge National Laboratory, Oak Ridge, TN 37831}

\date{\today}

\begin{abstract}
The magnetic and transport properties of Fe-deficient Fe$_5$GeTe$_2$ single crystals (\FGTx with $x\approx$0.3) were studied and the impact of thermal processing was explored.  Quenching crystals from the growth temperature has been previously shown to produce a metastable state that undergoes a strongly hysteretic first-order transition upon cooling below $\approx$ 100\,K.  The first-order transition impacts the magnetic properties, yielding an enhancement in the Curie temperature \TC from 270 to 310\,K.   In the present work, $T_{HT}\approx$550\,K has been identified as the temperature above which metastable crystals are obtained via quenching.  Diffraction experiments reveal a structural change at this temperature, and significant stacking disorder occurs when samples are slowly cooled through this $T$ range.   The transport properties are demonstrated to be similar regardless of the crystal's thermal history. The scattering of charge carriers appears to be dominated by moments fluctuating on the Fe(1) sublattice, which remain dynamic down to $\approx$ 100-120\,K.  Maxima in the magnetoresistance and anomalous Hall resistance are observed near 120\,K. The Hall and Seebeck coefficients are also impacted by magnetic ordering on the Fe(1) sublattice. The data suggest that both electrons and holes contribute to conduction above 120\,K, but that electrons dominate at lower $T$ when all of the Fe sublattices are magnetically ordered.  This study demonstrates a strong coupling of the magnetism and transport properties in \FGTx  and complements the previous results that demonstrated strong magnetoelastic coupling as the Fe(1) moments order. The published version of this manuscript is DOI:10.1103/PhysRevMaterials.3.104401 (2019).
\end{abstract}

\maketitle

\section{Introduction}

Van der Waals (vdW) bonded materials with quasi-2D crystal structures present many scientific opportunities for materials physics and can promote advanced functionality by combining complementary properties in van der Waals heterostructures.\cite{Geim2013}  Magnetically-active vdW materials are particularly attractive for complementing the electrical and optical properties of materials like graphene and transition metal dichalcogenides.\cite{Burch2018,Li2019} Indeed, several magnetic vdW materials have been exfoliated to the monolayer limit and magnetic order was found to persist, typically with a reduction in critical temperature relative to that of the bulk.\cite{Park2016,Duong2017,Burch2018,Li2019} Naturally, such pursuits promote a desire to identify vdW materials with magnetic order above room temperature.

Fe$_{5-x}$GeTe$_2$, sometimes referred to as \FGT\,, has recently emerged as an interesting vdW material.\cite{Stahl2018,ACSNano}  This phase is unique from Fe$_{3-x}$GeTe$_2$,\cite{Deiseroth2006,Chen2013,Verchenko2015,May2016,Liu2017,Wang2017,Yi2017,Nguyen2018} which has been  studied as a vdW ferromagnet that can be exfoliated to the monolayer limit.\cite{Fei2018,Deng2018}  In the bulk, Fe$_{5-x}$GeTe$_2$ has a higher Curie  temperature (260-310\,K) than Fe$_{3-x}$GeTe$_2$ (150-240\,K).\cite{Deiseroth2006,Chen2013,Verchenko2015,May2016,Stahl2018,ACSNano} For Fe$_{5-x}$GeTe$_2$, magnetic order has been demonstrated near room temperature in exfoliated flakes with thicknesses of 10\,nm.\cite{ACSNano}  In both materials, Fe vacancies likely play a key role in determining the physical properties.  Based on chemical analysis via wavelength dispersive spectroscopy, the vapor-grown crystals utilized in this study have a composition of Fe$_{4.7(2)}$GeTe$_2$.\cite{ACSNano}

\begin{figure}[ht!]%
\includegraphics[width=0.6\columnwidth]{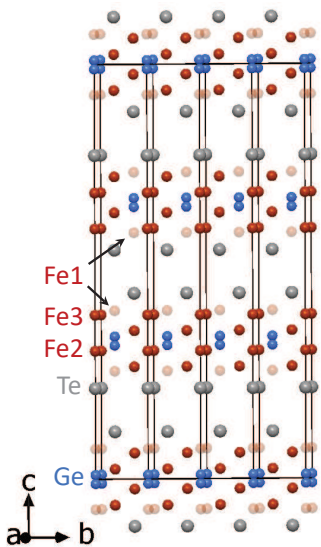}%
\caption{(color online). Schematic of \FGTx with atomic types labeled. Fe(1) and Ge were modeled as split sites; space group $R\bar{3}m$.  Atomic positions are (0,0,$z$) with Fe(1) $z$= 0.07305(17), Fe(2) $z$=0.30917(7), Fe(3) $z$=0.39597(9), Ge $z$=0.01074(10), and Te $z$=0.21896(4) from refinements at 220\,K, with $a$=4.0441(13)\,\AA\, and $c$=29.247(9)\,\AA\,.\cite{ACSNano}}%
\label{Structure}%
\end{figure}

Two similar crystal structures have been reported for Fe$_{5-x}$GeTe$_2$, with structures derived from single-crystal diffraction data collected using crystals that were synthesized differently.\cite{Stahl2018,ACSNano}  These structures with rhombohedral lattice centering have three \FGTx layers per unit cell, as shown in Fig.\ref{Structure}, and at complete occupancy the composition would reach Fe$_{5}$GeTe$_2$ if the vdW gaps remain empty.   Both reported structures possess vacancy-induced disorder and a Ge split site.  Our model, obtained using data collected on quenched crystals has a higher symmetry (space group $R\bar{3}m$, No. 166)\cite{ACSNano} than that reported by Stahl \textit{et al} (space group $R3m$, No. 160).\cite{Stahl2018}  As illustrated in Fig.\ref{Structure}, this model contains 3 Fe sites per unit cell.  The Fe(1) position is treated as a split-site and it can be occupied either above or below the neighboring Ge atom, or Fe(1) can be vacant.  Depending on occupation, the Ge atoms shift along the $c$-axis to maintain an appropriate bond distance (hence a split site for Ge).  In the lower symmetry model, the Fe(1) equivalent positions are not treated as split sites and are always `up' in a layer.  STEM imaging on vapor grown crystals observed different types of short range order associated with the Fe(1) site occupation, supporting the Fe(1) split site model for such crystals.\cite{ACSNano}    The lower symmetry model comes from data collected on crystals that were cooled naturally in the furnace and the higher symmetry model came from crystals that were vapor grown and quenched from 1023\,K into an ice-water bath.  The act of quenching was observed to greatly reduce the broadening of x-ray diffraction peaks,\cite{ACSNano} which were modeled by Stahl \textit{et al} as due to domains containing stacking faults.\cite{Stahl2018}  These prior results suggest a phase transition in \FGTx exists between 1023\,K and room temperature, making quenched crystals metastable.  Addressing the thermal stability of quenched crystals is one aspect of this manuscript.

\FGTx is essentially an easy-axis ferromagnet with moments preferring to orient along [001], though the magnetism is not particularly simple due to the presence of multiple Fe sublattices.  Depending on the thermal history, \TC ranges from 260 to 310\,K.  The compositional dependence of \TC has not been established. M\"ossbauer spectroscopy on a polycrystalline sample revealed that moments on the Fe(1) sublattice order below $\approx$100-120\,K while the majority of moments order at \TC.  Interestingly, crystals that are quenched from the growth temperature have \TC $\approx$270\,K, but upon cooling below $\approx$ 100\,K they undergo a `permanent' first-order transition to a state with enhanced magnetization for which \TC =310\,K.\cite{ACSNano}  This is reminiscent of relaxation effects that occur in FeMnP-based magnetocaloric materials upon their first thermal cycling.\cite{Tegus2013}  A reversible magnetoelastic effect was observed in polycrystalline \FGTx that is coupled to the ordering on the Fe(1) moments.\cite{ACSNano}   The dynamic behavior of the Fe(1) sublattice and the associated magnetoelastic effect appear to be dominant in \FGTx crystals regardless of their thermal histories. The intrinsic magnetoelastic effect is likely a driving force for the first-order transition in the metastable, quenched crystals.  After the first thermal cycle, the magnetization \MT of the quenched crystals are not thermally-hysteretic (for 2\,K $<T<$ 380\,K).  A magnetoelastic effect associated with the magnetic order of the Fe(1) sublattice is expected to persist even in the phase with enhanced \TC.  Non-quenched crystals (furnace cooled) also have non-hysteretic \MT but they are distinct from those of quenched crystals.  In all cases, the saturation moment at 2\,K is $\approx$2$\mu_B$/Fe and the Fe(1) sublattice remains (mostly) dynamic down to approximately 100\,K.

In this work, we compare the magneto-transport and magnetic properties among \FGTx crystals ($x \approx$0.3) with the thermal histories described above. The temperature dependence of the resistivity and the behaviors of the Hall effect and magnetoresistance are essentially independent of the processing conditions, indicating these that properties are not particularly sensitive to stacking faults.  A maximum is observed in the magnetoresistance and in the anomalous Hall effect near 120\,K, close to where the Fe(1) sublattice orders magnetically. The transport data suggest that both electrons and holes contribute above 120\,K, but electrons appear to dominate transport at lower $T$.  X-ray diffraction experiments up to 693\,K were performed and reveal a structural transition near 550\,K. Magnetization measurements confirm that crystals quenched from above this temperature are metastable.

\section{Methods}

Single crystals of \FGTx were grown in the presence of iodine under nearly-isothermal conditions (muffle furnace).\cite{ACSNano}  Nominal compositions ranging from Fe$_{4.7}$GeTe$_2$ to Fe$_{6.0}$GeTe$_2$ were explored and similar magnetic properties were observed for the resulting crystals.  Previous wavelength dispersive spectroscopy experiments suggested a composition of Fe$_{4.7(2)}$GeTe$_2$ for these crystals, while refinement of single crystal x-ray diffraction data yielded Fe$_{4.87(2)}$GeTe$_2$. We utilize the composition based on chemical analysis as the identifier for these crystals.  A few growth temperatures were also explored and the only processing condition that impacted the observed magnetic behavior was the cooling rate.  This implies that the growths occur at a fairly specific composition under these conditions.

The crystal growths occurred in vacuum sealed, argon-purged silica ampoules. The ampoules were generally 150\,mm long with an inner diameter of 22\,mm and an outer diameter of 25\,mm; 0.1-0.2\,g of iodine was utilized in such growths.  Some growths were allowed to cool naturally in the furnace and these batches produced `non-quenched crystals' while other growths were quenched into ice-water baths from 1023\,K.    For quenched crystals, iodine was rinsed from the surface using acetone and/or alcohol.  When crystal growth was performed in an intentional temperature gradient, the binary Fe$_{1+x}$Te phase sometimes formed at the cold end and the \FGTx phase grew at the hot end on the source material. Crystal growths were performed starting with elemental Fe (99.98\%, granules), Ge (99.999\%, pieces) and Te (99.9999\%, shot) with metals basis purity listed.  A polycrystalline sample was also studied, which was synthesized using elemental powders that were ground together in a glovebox and the product was quenched from the reaction temperature of 973\,K.  Some of this polycrystalline sample was previously utilized for neutron powder diffraction and M\"ossbauer spectroscopy.\cite{ACSNano}  For annealing studies, the samples were sealed in silica ampoules with argon exchange gas and quenched into an ice-water bath from the annealing temperature.

X-ray powder diffraction data were collected on as-quenched samples without grinding to minimize the impact of preferred orientation and samples were spun during data collection.  Room-temperature data were collected with a PANalytical X'Pert Pro MPD utilizing a Cu K$\alpha_1$ ($\lambda$=1.5406\,\AA) incident beam monochromator.  

High temperature x-ray diffraction data were collected on a PANalytical Empyrian X-ray diffractometer with Cu K$\alpha$ radiation. An Anton Paar XRK900 heating stage employing automatic variable temperature height adjustment was used to obtain data at each temperature, with helium flowing over the sample to minimize oxidation.  Experiments performed on a single crystal utilized an average ramp rate of 2\,K/min (including  time spent as isothermal measurements every 20\,K).  For the polycrystalline experiments,  an initial 6\,h data collection and purging were performed at 298\,K.  The sample was then heated at 5\,K/min to 373\,K and subsequent ramping between data collections at 373 and 693\,K were performed at 2\,K/min.  For the first two cycles, the measurement times at 373\,K and 693\,K were six hours each, and thereafter data collections of 3\,h were employed.  These measurements on powders were perfomed in 0.5\,h scans and the patterns evolved during the first 1-2\,h at 373\,K on the warming cycle. The PANalytical software HighScore was utilized to sum consecutive scans and strip the diffraction from K$\alpha_2$ radiation, as well as to obtain $d$-spacings of individual reflections.  Le Bail fitting and Rietveld refinements were performed using the program FullProf.\cite{FullProf}

Physical property measurements were performed in Quantum Design Physical and Magnetic Property Measurement Systems.  The properties of non-quenched crystals are not thermally hysteretic, at least in the range 2 $< T <$ 380\,K,\cite{ACSNano} while the magnetic properties of the quenched crystals are impacted by a first-order transition that occurs during the first cooling cycle to cryogenic temperatures.  Data were collected upon cooling in an applied field unless otherwise noted.  

\section{Results and Discussion}

We begin by stating the nomenclature utilized to differentiate the thermal histories investigated.  The crystals are non-quenched (NQ), quenched but not cooled below 200\,K (Q-HT), and quenched plus thermally cycled (Q-C) to well below 100\,K to induce the first order transition that enhances \TC. The identifier Q-HT is meant to convey that the crystals are quenched with a high-$T$ phase and thus metastable.  We also examined some polycrystalline specimens.  The powders do not possess a first order transition even though they are quenched from the reaction temperature.   The lack of thermal hysteresis in the magnetization of quenched powders is perhaps due to the inability to rapidly quench powders (lightly sintered and large masses), the lack of microstrain in powders due to the presence of additional defects, or some subtle differences in composition and/or short range orderings.  A reversible magnetoelastic effect has been observed in the powder samples near 100\,K via neutron powder diffraction.\cite{ACSNano} This coupling of magnetism and the lattice involves the magnetic order on the Fe(1) sublattice, and the behavior is expected to be present in all versions of \FGTx inspected here (even Q-C).

\subsection{Metastability in Quenched \FGTx Crystals}

\begin{figure}[ht!]%
\includegraphics[width=\columnwidth]{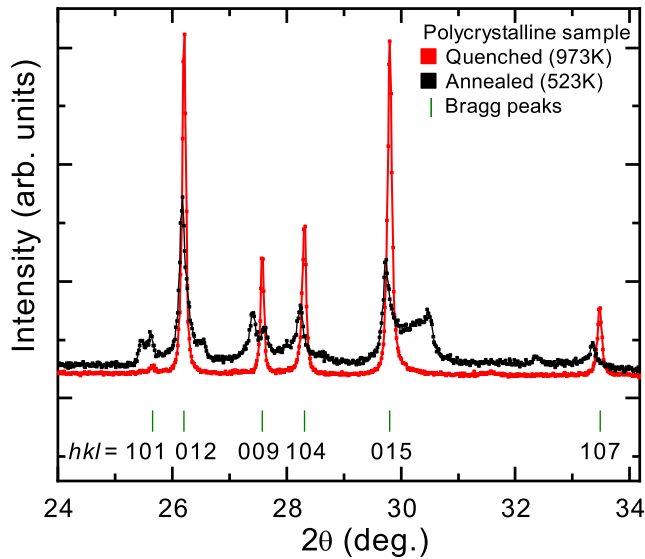}%
\caption{(color online). X-ray powder diffraction data for a polycrystalline sample of nominal composition Fe$_{4.7}$GeTe$_2$ that was quenched (red markers) and annealed at 523\,K for 257\,h (black markers).  Bragg peaks are indexed and marked by the green ticks based on refinement of the data for the quenched sample.}%
\label{pxrd}%
\end{figure}

The impact of annealing on the x-ray powder diffraction data is shown in Fig.\,\ref{pxrd} for a polycrystalline sample of nominal composition Fe$_{4.7}$GeTe$_2$.  When quenched from the reaction temperature of 973\,K, the diffraction data are characterized by sharp Bragg peaks.  Rietveld refinement of the data yield $a$=4.0348(2)\AA\, and $c$=29.088(1)\AA.  We note that the single crystals and the powders reported previously had slightly larger $c$-axis lattice parameters and it is difficult to know if this is an effect of composition or thermal history.\cite{Stahl2018,ACSNano}  Figure\,\ref{pxrd} shows that annealing the quenched powder at 523\,K for 257\,h (followed by quenching)  causes the diffraction data to have broadened $h0l$ reflections, as well as the emergence of some small peaks that might be due to impurities.  The two curves in Fig.\,\ref{pxrd} were generated using two different potions of a 12\,g polycrystalline reaction. Similar diffraction data are observed for polycrystalline samples that are allowed to cool to 300\,K in the furnace after the high temperature reaction.\cite{Stahl2018}  This reveals that some structural modification (or degredation) occurs between 523 and 973\,K. Upon annealing at 523\,K, both lattice parameters increase, though the data are not well modeled due to the broadening and peak shifting that is likely caused by stacking faults.\cite{Stahl2018}  We also infer an increase in the $c$-axis lattice parameter upon annealing quenched crystals at 523\,K by inspecting diffraction data collected from the facets of quenched/annealed single crystals.

\begin{figure*}[ht!]%
\includegraphics[width=1.7\columnwidth]{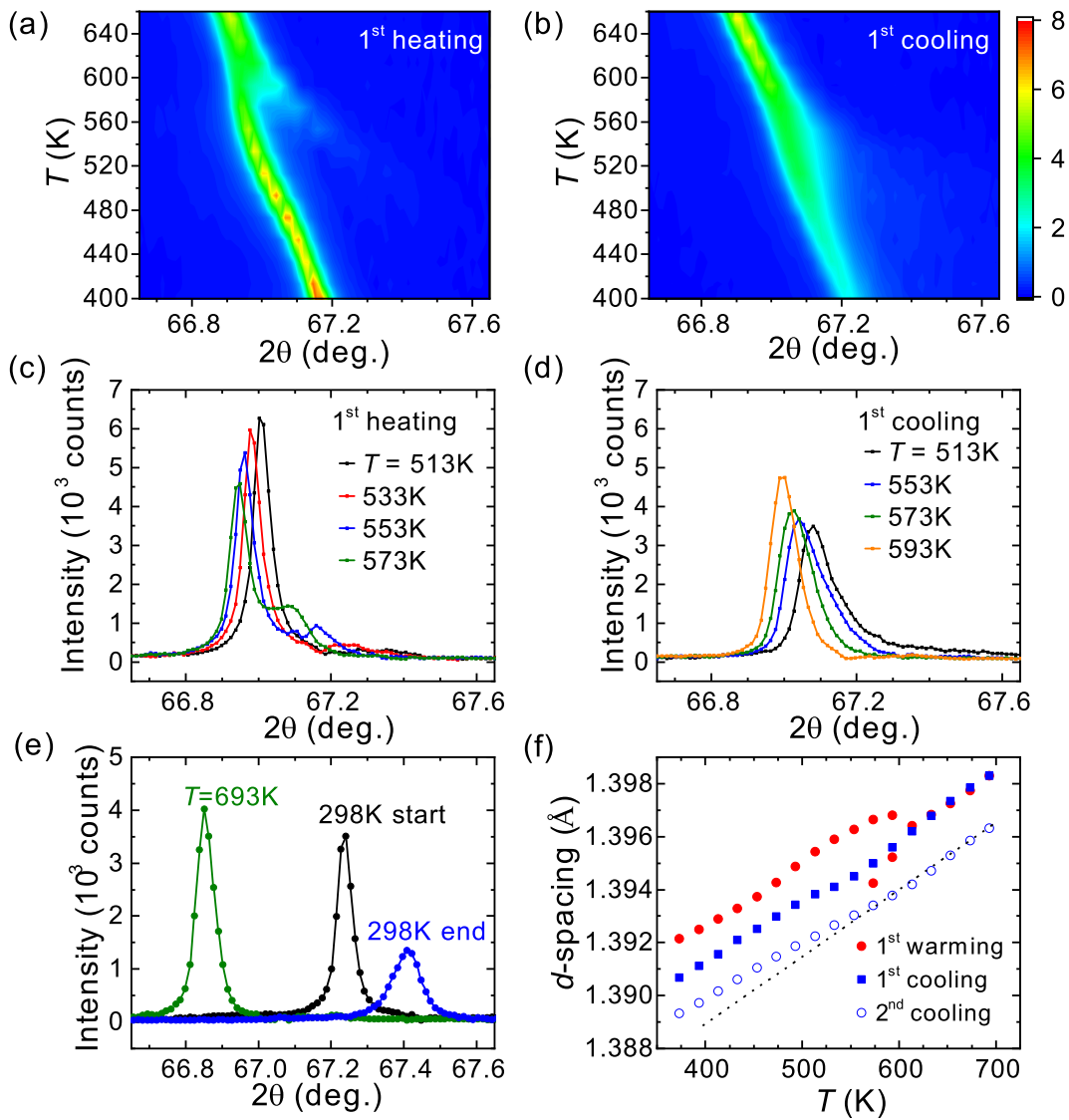}%
\caption{(color online). Temperature-dependent x-ray diffraction data for 0\,0\,21 Bragg reflection from the facet of a Q-HT single crystal. Intensity upon (a,c) first heating and (b,d) first cooling.  (e) Comparison of intensity in the metastable quenched state (`298K start'), at the maximum temperature of 693\,K, and at the end of the experiment (`298K end'). (f) Fitted d-spacing of the 0\,0\,21 reflection upon thermal cycling; the dashed line is a guide to the eye for expansion in the high $T$ phase. The color maps in (a,b) share the same intensity scale that is shown beside panel (b) with units of 10$^3$ counts.}%
\label{HTxrd}%
\end{figure*} 

\begin{figure}[ht!]%
\includegraphics[width=\columnwidth]{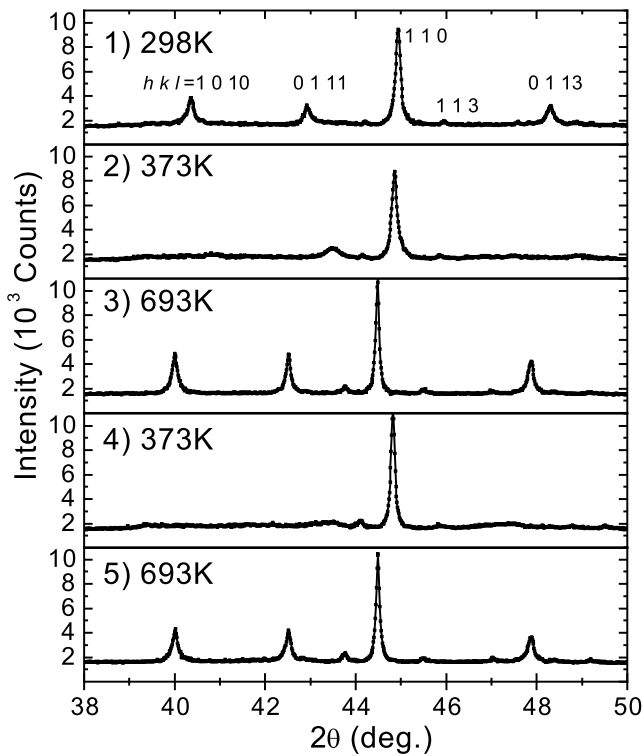}%
\caption{(color online). X-ray diffraction data for a quenched polycrystalline sample of \FGTx as a function of temperature.  Bragg reflections are indexed in the top panel. The numbers in each panel indicate the step number for the sequence utilized, which contained two warming cycles to 693\,K.  Upon first heating to 373\,K, small changes in the data were noted during the first 1-2\,h, and thus the data shown in panels 2 and 3 were collected after a 3\,h wait at each temperature; all patterns shown are for the same total collection time.}%
\label{powderHT}%
\end{figure} 

\begin{figure}[ht!]%
\includegraphics[width=\columnwidth]{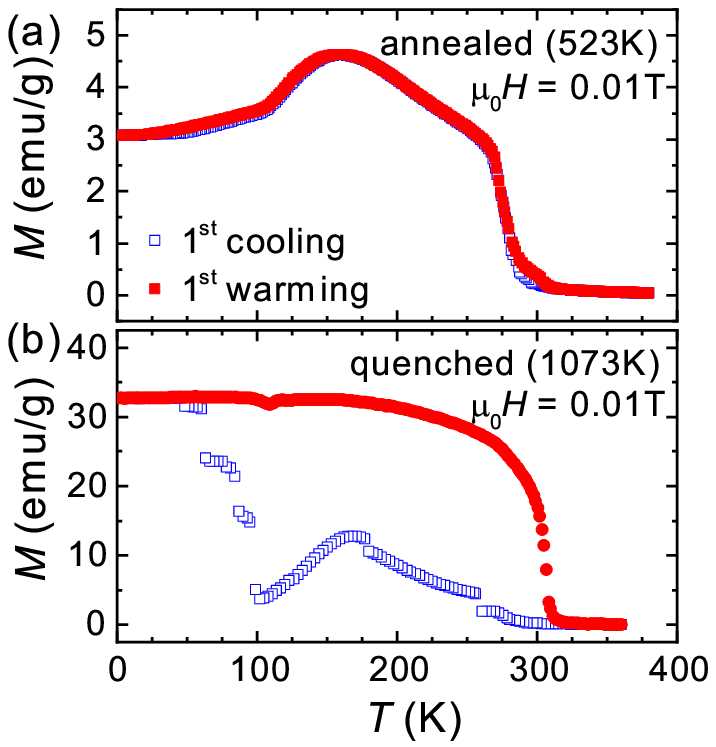}%
\caption{(color online). Magnetization of (a) single crystal that was annealed at 523\,K; data for field applied within the basal plane, and (b) crystals that were quenched from 1023\,K with hysteresis and an enhancement in \TC observed.  Data in (b) were obtained as part of the measurement sequence presented in Fig.\,\ref{QuenchedCyling}; \TC=310\,K remains for subsequent measurements.}%
\label{MT_anneal}%
\end{figure} 

To provide more evidence for the structural change that is implied by the diffraction data in Fig.\,\ref{pxrd}, we performed a temperature-dependent x-ray diffraction study on a quenched single crystal.  Results are shown in Fig.\,\ref{HTxrd} for the 0\,0\,21 Bragg reflection, which is representative of the behavior of all the 00$l$ reflections.  The behavior of the $l$=21 reflection upon the first heating and cooling is shown in Fig.\,\ref{HTxrd}(a-d).  Upon warming, the reflection broadens towards higher-angles and a second peak is  observed starting near 530\,K (Fig.\,\ref{HTxrd}(c)). Similar behavior was observed upon warming the second time as well.  This may suggest a coexistence of structural phases, which is consistent with a first-order structural change, though clear coexistence is not observed upon cooling (Fig.\,\ref{HTxrd}(d)) where the transition looks more second-order.  For a given 00$l$ reflection, a single peak is clearly observed above $\approx$ 600\,K.  Upon cooling, the broadening returns (Fig.\,\ref{HTxrd}(b)) and the intensity decreases (Fig.\,\ref{HTxrd}(d)).  Based upon these results, we define the transition temperature as $T_{HT}\approx$550\,K.  After thermal cycling, the diffraction peaks are broadened as compared to the first scan on the quenched crystal (Fig.\,\ref{HTxrd}(e)), consistent with the behavior of the polycrystalline samples.  After thermal cycling, the integrated area for the highest intensity reflection (0\,0\,9) is within 4\% of its starting value.  The sharpest reflections and highest peak intensities are observed in the high $T$ state.  After the high-temperature x-ray diffraction experiment, the magnetization \MT data obtained for this crystal were consistent with the data obtained on non-quenched crystals.

An irreversible change in the $c$-axis lattice spacing occurred during the high-temperature diffraction experiment. This is qualitatively illustrated by the position of the reflection in Fig.\,\ref{HTxrd}(e).  The shift in 2$\theta$ at 298\,K between the first scan and the last scan corresponds to a decrease in the $c$-axis lattice parameter after warming above the transition and cooling back through $T_{HT}$.  Results for the $c$-axis parameter obtained from Le Bail fittings are shown in Table \,\ref{facet}; data containing 00$l$ reflections between 5 and 100 degrees 2$\theta$ were fit.  The results are consistent with our previous STEM experiments that revealed a smaller $c$-axis parameter for non-quenched crystals in comparison to that observed in quenched crystals.  Recall this crystal is starting the experiment in a quenched and metastable state.  

\begin{table}[h]
\caption{$c$-axis lattice parameters from fitting temperature-dependent diffraction data off a single crystal facet with heating/cooling sequence indicated by scan number.}
\begin{tabular}[c]{|c|c|c|}
\hline
Scan No. &  $T$ (K) & $c$ ($\AA$) \\
\hline
1 & 298 & 29.2370(10) \\
2 & 373 & 29.2549(6)  \\
18 & 639 &  29.3876(8) \\
34 & 373 & 29.209(2)  \\
50 & 639 &  29.354(2)  \\
67 & 298 & 29.177(1)  \\
\hline
\end{tabular}
\label{facet}
\end{table}

The changes in lattice spacing that occur while warming/cooling through the structural change can be viewed by tracking the $d$-spacing of a given Bragg reflection, as shown in Fig.\,\ref{HTxrd}(f).  The $d$-spacing, which is directly proportional to the layer spacing in the crystal structure, decreased upon warming into the high-$T$ structure during both the first and second warming cycles.  The changes upon cooling are more subtle, but clearly noticeable when the effect of thermal expansion is also considered (the dotted line in Fig.\ref{HTxrd}(f) is meant to reveal this effect).  The fact that the $d$-spacing decreased again on the second thermal cycle may suggest a change in composition is occurring across the transition, but these details are not accessible with the current data and the trends could be caused by continuous changes in short range order or locations of atoms within the cell (kinetically limited).  The data in Fig.\,\ref{pxrd} for powders annealed for over 10\,d suggest that perhaps some impurities are formed but that the phase remains after a long-duration at a $T$ just below $T_{HT}$. 

A temperature-dependent diffraction experiment was also completed using a polycrystalline sample and the results are shown in Fig.\,\ref{powderHT}.  A quenched powder with well-defined Bragg peaks was utilized (Fig.\,\ref{powderHT} panel 1); a small amount of unknown impurity was detected in this portion of sample (peak at 2$\theta$=44.2$^{\circ}$).  The experiment revealed that transformation to a state with significant stacking disorder happens rapidly at 373\,K in the powder sample.  This is evidenced by the significant suppression of intensity for $h$0$l$ reflections in panel 2 of Fig.\,\ref{powderHT}.  However, the 110 Bragg reflection remains sharp at 373\,K (even after cooling from high temperature, see panel 4). This indicates that the in-plane structure is not strongly affected by the structural change.  These results are generally consistent with those in Fig.\,\ref{pxrd} for the powder annealed at 523\,K.  The \textit{in situ} diffraction experiments resulted in greater stacking disorder, however, as evidenced by a stronger suppression of $h$0$l$ reflections.  The results in Fig.\,\ref{powderHT} were obtained using a portion of the quenched polycrystalline sample that was also utilized to perform the annealing study (Fig.\,\ref{pxrd}).  The powder looked slightly oxidized (brownish) after the \textit{in situ} measurements, which may relate to the increase in impurity content.  The measurements were performed in the following sequence: 1) quenched powder at 298\,K, 2) 373\,K, 3) 693\,K, 4) 373\,K, 5) 693\,K, 6) 373\,K, 7) 298\,K.  Data for steps 1-5 are shown in Fig.\,\ref{powderHT} with panels labeled by the step number.  Data collected in steps 6 and 7 are not shown and are consistent with those in steps 2 and 4 (broadened $h$0$l$ peaks).  After this experiment, despite some apparent oxidation, the temperature-dependence of the magnetization of the powder is consistent with that of the furnace cooled single crystals. 

These results do not directly probe the kinetics of the transformation of the metastable phase below $T_{HT}$ but some information is gained.  The polycrystalline sample shows the impacts of disorder (transformation) in diffraction data rapidly at 373\,K.  The data in panel 2 of Fig.\,\ref{powderHT} were collected after 3\,h at 373\,K because the diffraction pattern was clearly evolving during the first 1-2\,h of data collection.   The data obtained for Q-HT single crystals do not show broadening until over 500\,K, although the 00$l$ reflections are not especially sensitive to the stacking disorder that evidences the transition (and duration at 373\,K was different for the two experiments).  It is certainly possible that the kinetics are different in the crystals and the powders, and it is worth noting again that the polycrystalline sample  had slightly different lattice parameters and the magnetic properties were not thermally hysteretic at low $T$.  It is also worth highlighting that the experiment does not probe the stability of the Q-C phase with the highest \TC.  This phase is probably more stable than the Q-HT phase, since the first-order transition relieves strain along [001] and induces stacking faults.\cite{ACSNano}

The high-temperature diffraction data clearly demonstrate a structural transition occurs near 550\,K.  Significant stacking disorder occurs upon cooling into the low $T$ state ($T <$ 550\,K). The in-plane structure and average layer stacking remain coherent, although a reduction in domain size may occur.  The transition appears to be reversible because both powder measurements at 693\,K produced patterns with sharp diffraction peaks at similar positions.  Also, the total integrated intensity of the 009 reflection does not change significantly during the single crystal experiment.  These results imply that a simple decomposition is not likely to be the source for the structural change.  Nanoscale probes would be necessary to determine if the transition relates to a eutectoid reaction that is kinetically or chemically hindered.

Stahl \textit{et al} found that a combination of faultless and faulted domains best described the diffraction data of \FGTx powders containing significant amounts of stacking disorder.\cite{Stahl2018}  In light of the current results, the simulations by Stahl \textit{et al} may have revealed that in some regions of the material the structural change upon cooling through $T_{HT}$ is incomplete (kinetics) or incompatible with the local composition or the local short range order.    The main impact on the average structure is the stacking disorder, which implies that a stacking-related transition occurs.  Several types of symmetry changes could be associated with such a stacking transition, including the shift to a monoclinic cell as in vdW CrI$_3$ and CrCl$_3$.\cite{McGuire2015,McGuire2017}  In those systems, the main effect is a change in layer spacing with a larger layer spacing observed in the monoclinic phases that are the high $T$ states.   Other types of stacking-related transitions are also possible, such as transitioning to one layer per unit cell or from the rhombohedral ABC stacking to a primitive AAA stacking.  For all of these cases, intrinsic disorder (structural or chemical) could hinder a coherent structural transition and result in a crystal with significant stacking disorder. It is worth emphasizing that the structural model in Fig.\,\ref{Structure} is an average structure for the high $T$ state because quenched (Q-HT) crystals were utilized for the single crystal x-ray diffraction data collection.  The model obtained from the metastable quenched crystal has a higher symmetry and an associated increase in disorder (entropy) as compared to the model obtained by Stahl \textit{et al} on naturally cooled samples.\cite{Stahl2018}  Both models utilize rhombohedral lattice centering with ABC stacking. 

Another explanation for the change in layer spacing, which could occur simultaneously with a change in space group, would be the existence of a small amount of Fe residing within the gap at high temperature.  This may increase the bonding between the layers and reduce the layer spacing, which is what we observe upon warming above $T_{HT}$. The presence of Fe in the vdW gap at high $T$ would be entropically favored and could lead to a greater importance of kinetics in our \textit{in situ} experiments.  Fe atoms leaving the van der Waals gap and entering the metallic slabs (below $T_{HT}$) would likely increase the $a$ lattice parameter, consistent with our annealing studies.  Our original single crystal diffraction data on quenched crystals did not clearly reveal electron density within the van der Waals gap. However, a small amount of disordered Fe within the gap could be difficult to detect.  The movement of copper into and out of the van der Waals gap was observed in CuInP$_2$S$_6$, where copper leaves the metallic slabs and enters the vdW gap above $T\approx$315\,K.\cite{Maisonneuve1997} Movement of copper ions is fairly common in materials, though the defects in \FGTx may promote the movement of Fe.  Due to the significant amount of stacking disorder,  quantitative refinements of the diffraction data collected below 550\,K were not possible.  Utilizing single crystal diffraction (four circle) to examine the thermal displacement parameters upon cooling towards $T_{HT}$ may prove especially insightful in isolating the microscopic origins of the phase transition.  The extent to which the behavior around $T_{HT}$ varies with Fe content could also be particularly informative, if the Fe content can be manipulated.  Magnetization measurements after annealing at 523\,K did not reveal a soft ferromagnetic component in \MH data at 350\,K, and thus there is currently no evidence that any significant amount of Fe is being ejected from the crystal below $T_{HT}$.  

With regard to \FGTx, it may be useful to consider the lattice trends in Fe$_{3-x}$GeTe$_2$, another metallic vdW material with atomic disorder.  In Fe$_{3-x}$GeTe$_2$, the $c$-axis parameter decreases while $a$ expands upon the filling of Fe vacancies in the metallic slab.\cite{May2016} Such a decrease in $c/a$ can be viewed as typical for a quasi-2D material where the layer spacing is dominated by the bond distance between the large anions (which are forced apart when the intralayer distances increase, causing a compression along $c$ to compensate).  The scenario would be different in \FGTx if the total Fe content is fixed and Fe moves in and out of the vdW gap.  The role of short range orders may be critical in \FGTx, necessitating the use of real-space local probes such as scanning transmission electron microscope (STEM).  In this regard, it is worth noting that previous STEM experiments revealed two types of short rage order.  One of them produced alternating slab thicknesses, and could thus drive incoherence along [001] (see supporting information for Ref.\,\citenum{ACSNano}).

These results demonstrate that \FGTx crystallizes into a high-$T$ state and a structural transition that results in significant stacking disorder occurs upon cooling through $T_{HT}\approx$550\,K. By quenching from $T > T_{HT}$, a metastable crystal is thus obtained and we identify such crystals by the label Q-HT.  When the metastable Q-HT crystal is cooled sufficiently low in temperature ($\approx$100\,K), it undergoes a transition to a new state that we identify using the label Q-C.  The Q-HT and Q-C states are magnetically unique from one another and from the state that is obtained by cooling slowly in the furnace (non-quenched, label NQ).  The Q-C and NQ crystals have a large concentration of stacking faults that precluded us from solving the crystal structures via single crystal x-ray diffraction.

The impact of thermal history on the magnetization of quenched single crystals is introduced in Fig.\,\ref{MT_anneal}.  Figure\,\ref{MT_anneal}(a) plots the temperature-dependent magnetization of a single crystal that was annealed at 523\,K for 257\,h.  The crystal was originally quenched from its growth temperature of 1023\,K and was also quenched from the annealing temperature of 523\,K.  The observed \MT is similar to what is seen in NQ single crystals that are allowed to cool naturally in the furnace (see below).  The \MT data for a quenched crystal are shown in Fig.\,\ref{MT_anneal}(b). Quenched crystals display a strong thermal hysteresis upon the first cooling below $\approx$100\,K, with enhanced magnetization observed below 100\,K and upon warming.  Remarkably, the first-order phase transition near 100\,K impacts the magnetization near 300\,K with an enhancement in \TC that is non-hysteretic in subsequent thermal cycling to cryogenic temperatures.

\begin{figure}[ht!]%
\includegraphics[width=\columnwidth]{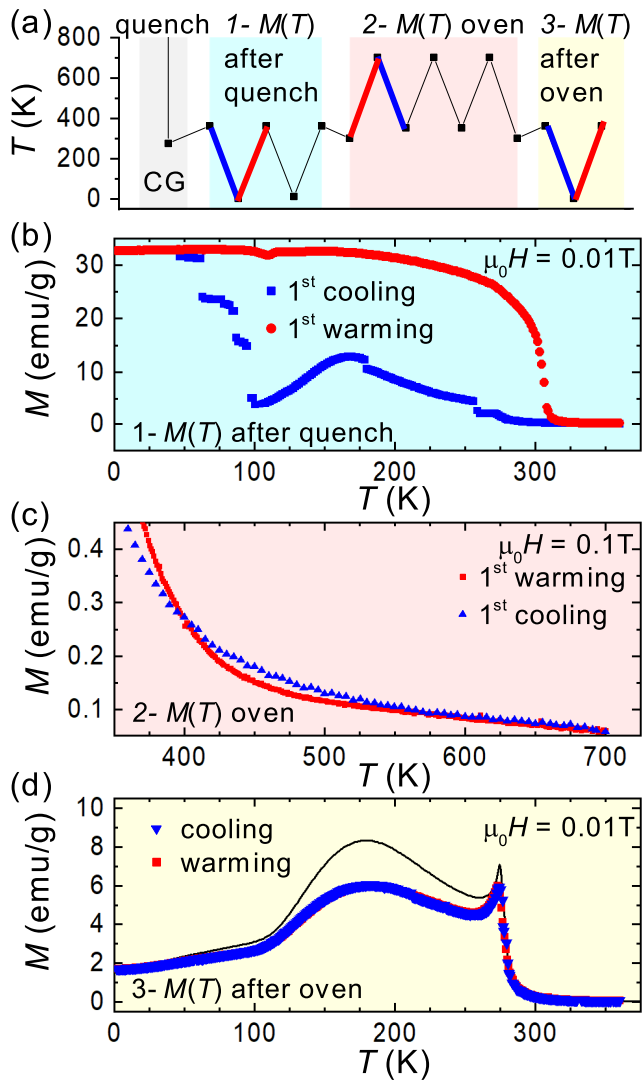}%
\caption{(color online). (a) Sequence of thermal treatment and magnetization measurements starting with quenching from crystal growth (CG) temperature of 1023\,K.  Red and blue lines indicate segments during which data in panels (b-d) were collected.  (b) Data obtained during first cooling/warming cycle down to 2\,K with hysteresis. (c) High temperature magnetization revealing collapse of warming/cooling data near 570\,K. (d) Low-$T$ magnetization data after heating to 700\,K, with comparison to data for a non-quenched crystal (solid line, $H \perp c$).}%
\label{QuenchedCyling}%
\end{figure} 

The thermal stability of quenched and non-quenched crystals was examined by \MT measurements during thermal cycling. The results shown in Fig.\,\ref{QuenchedCyling} demonstrate the existence of the different states of \FGTx crystals in the order of Q-HT, Q-C, and finally NQ. The thermal history for this experiment is illustrated in Fig.\ref{QuenchedCyling}(a).  The experiment began with quenching from the crystal growth (CG) temperature of 1023\,K. In sequence, \MT data were obtained while crystals were cooled to 2\,K (1 - light blue area), heated to 700\,K (2 - light red area) and then cooled to 2\,K (3 - yellow area).  The data shown in Fig.\,\ref{QuenchedCyling}(b-d) correspond to the blue (cooling) and red (warming) curves outlined in Fig.\,\ref{QuenchedCyling}(a).  The data in Fig.\ref{QuenchedCyling}(b-d) were collected on a single set of non-orientated crystals using a silica sample holder.  While the crystals are not strictly oriented, the largest field projection was for $H \perp c$ and the crystals are free to rotate somewhat; these plate-like crystals align $H\perp c$ when placed on a permanent magnet at room temperature.  Thus, it is reasonable to compare the data in Fig.\ref{QuenchedCyling} to results for \MT with $H \perp c$ (at least for $T<350$\,K), such as in Figs.\,\ref{MT_anneal}(a),\ref{MT_Compare},\ref{M_Compare} and Ref.\citenum{ACSNano}.

The first-order transition between Q-HT and Q-C is demonstrated in Fig.\ref{QuenchedCyling}(b) (repeated in Fig.\,\ref{MT_anneal}(b)).  The transition occurs at $\approx$100\,K, where a sharp increase in the magnetization is observed upon the first cooling cycle of Q-HT crystals.  In the \MT data, the transition is very sharp for small applied fields and the enhancement in \TC is very apparent.  If not cooled to near the transition at 100\,K, the magnetization of the quenched Q-HT crystals is reversible with a Curie temperature similar to that in NQ crystals (\TC$\approx$270\,K).

Magnetization data collected while warming to 700\,K are shown in Fig.\,\ref{QuenchedCyling}(c).  These data are collected after those in Fig.\,\ref{QuenchedCyling}(b), and thus the data for warming correspond to the Q-C state with \TC = 310\,K.  Heating and cooling rates during the magnetization measurements are relatively slow, ranging from 2-4\,K/min; isothermal sections were not utilized.  Thus, after the measurement in the oven the crystals are expected to behave like the non-quenched crystals.  This was indeed found to be the case, as illustrated in Fig.\,\ref{QuenchedCyling}(d) by the lack of hysteresis and the overall shape of \MT upon measuring from 360 to 2 to 360\,K . Magnetization data for an NQ crystal are shown as the solid curve in Fig.\,\ref{QuenchedCyling}(d) for comparison ($H \perp c$).

In Fig.\,\ref{QuenchedCyling}(c), the warming and cooling curves are converged above $\approx$570\,K.  The diffraction data shown in Fig.\,\ref{HTxrd} were collected more slowly than the magnetization data in Fig.\,\ref{QuenchedCyling}(c), and the diffraction data reveal the onset of a structural change as low as perhaps 530\,K for the metastable crystal.  However, the diffraction data were collected using a Q-HT crystal (not Q-C crystal). Upon cooling, the single crystal diffraction peaks start broadening around 560-580\,K, consistent with the magnetization results.  The change in \MT near $T_{HT}$ is negligible upon cooling, likely due to the fact that the dominant structure within each \FGTx slab remains intact across the transition according to the sharpness of the 110 and related reflections. Thus, the response in the paramagnetic region is not expected to be significant.

Additional measurements were performed to further verify that metastable crystals are obtained by quenching from above $T_{HT}\approx$550\,K, but the data are not shown for simplicity.  Crystals annealed at 573\,K (and quenched) displayed hysteretic \MT behavior consistent with a Q-HT to Q-C transition (similar to Fig.\,\ref{QuenchedCyling}(b)).  In addition, magnetization measurements on NQ crystals up to 700\,K demonstrated a convergence of heating/cooling data above $\approx$575\,K.

The \MT data suggest that the Q-C state of \FGTx crystals can be stable up to at most 570\,K.  While powders and crystals may present different kinetics for the transformation, the quenched powder samples show signs of transfomation to a state with significant stacking disorder by 373\,K.  However, magnetization measurements show that the Curie temperature of Q-C crystals is stable and reversible in the range 2-380\,K.  This suggests that the first-order transition from Q-HT to Q-C, which relieves lattice strain,\cite{ACSNano} results in a more stable state that can perhaps survive up to the transition at $T_{HT}$. However, the rapid rate of magnetization measurements could skew the results and a more detailed investigation of the kinetics is necessary.

\subsection{Magnetization}

\begin{figure}[ht!]%
\includegraphics[width=\columnwidth]{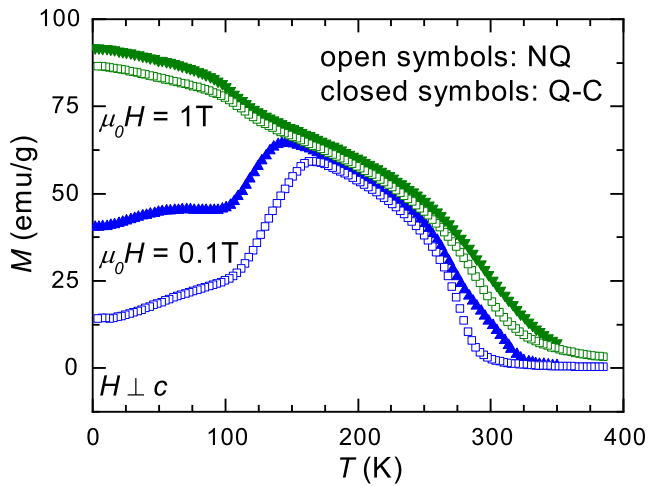}%
\caption{(color online). Comparison of temperature-dependent magnetization in non-quenched (NQ) and thermally-cyled quenched (Q-C) single crystals of \FGTx\, for $H \perp c$.  Data sets for two different applied fields are labeled accordingly; data collected upon cooling in an applied field.}%
\label{MT_Compare}%
\end{figure}

This section compares the magnetization $M$ in non-quenched (NC) to quenched plus thermally-cycled (Q-C) crystals.  The Q-C crystals have \TC=310\,K that is coupled to the first-order transition observed upon first cooling Q-HT crystals below 100\,K, as discussed in the previous section.  The non-quenched (furnace-cooled) crystals have \TC =275\,K.  As demonstrated by Fig.\,\ref{MT_Compare}, the  main differences in \MT are observed above $\approx$250\,K.  For NQ and Q-C crystals in the temperature ranges inspected in this section, the \MT data are not thermally hysteretic.

As shown in Fig.\,\ref{MT_Compare}, the existence of several characteristic features in \MT suggests that the magnetism is not necessarily simple. The magnetization in small magnetic fields is characterized by a sharp onset near \TC, followed by an increase in $M$ upon cooling towards a broad cusp.  The temperature at which the  cusp occurs depends on the orientation and magnitude of the applied field.\cite{ACSNano}  A prominent feature is observed near 100-120\,K with a reduced $M$ for cooling in small applied fields and a small enhancement in the magnetization for large applied fields.  As was demonstrated in Ref.\,\citenum{ACSNano}, the majority of the Fe(1) sublattice is magnetically disordered above $\approx$ 100-120\,K.  At 1.5\,K, when Fe(1) is ordered, neutron diffraction data were modeled equally well by a ferrimagnetic model or a ferromagnetic model. However, at 160\,K, the neutron powder diffraction data suggested ferromagnetic order.\cite{ACSNano}  Together with the maximum in \MT for fields less than the saturation field ($\approx$ 1\,T), these results may suggest that Fe(1) moments are not aligned with the others, either due to spin canting or anti-parallel alignment (ferrimagnet) for $H\approx 0$.

\begin{figure*}[ht!]%
\includegraphics[width=1.8\columnwidth]{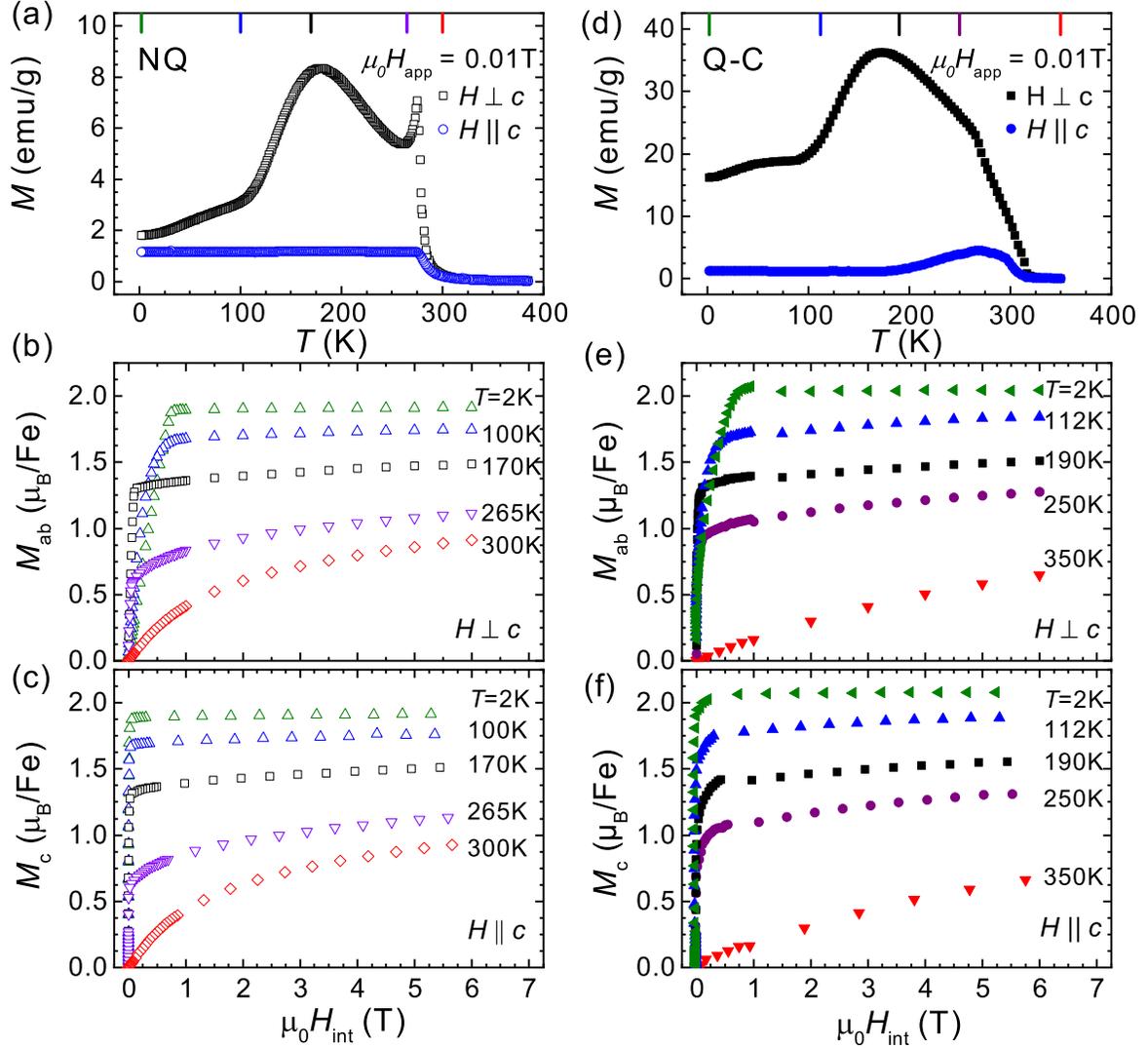}%
\caption{(color online). Anisotropic magnetization data for single crystals of \FGTx ($x\approx$0.3) comparing behavior for thermal histories non-quenched (NQ panels a,b,c) and quenched plus previously cooled to 2\,K (Q-C, panels d,e,f).  The top panels (a,d) demonstrate temperature-dependent magnetization for fixed applied field, and the ticks marks on the upper $x$-axis indicate the temperatures of the isothermal magnetization data shown in panels (b,c,e,f).   For $H \parallel c$ the demagnetization factor has been estimated as $N$=0.9 (see text), resulting in the plotted internal field $H_{\textrm{int}}$ being less than the applied field $H_{\textrm{app}}$.}%
\label{M_Compare}%
\end{figure*}

The anisotropic magnetization data are presented in Fig.\,\ref{M_Compare}. In the upper panels, Fig.\,\ref{M_Compare}(a,d), \MT are shown for an applied field of 0.01\,T (data collected upon cooling). For this relatively small applied field, a strong cusp is observed near 275\,K in \MT of the NQ crystal for $H \perp c$, while the data for Q-C crystals display a shoulder at this temperature.  The cusp may be a particularly interesting feature, perhaps relating to some complex spin texture or domain structure,\cite{Onose2005,Chattopadhyay2009,Ghirmire2013} though canting or the onset of ferrimagnetism could also cause such behavior.  The low-field results also differ in the extent to which the data for $H \parallel c$ change with $T$, with slightly more temperature dependence observed for the Q-C crystals (likely due to the sublattice component with \TC=310\,K). 

As shown in Fig.\,\ref{M_Compare}(a,d), the induced moment is larger for fields applied within the basal plane compared to when the field is applied along the $c$-axis. However, demagnetization effects impact the apparent anisotropy and \MT since the total field (external plus demagnetizing) is changing significantly as a function of temperature. Due to this complication, examining anisotropy via \MT or a susceptibility-like quantity as $M/H$  is not particularly useful or valid. 

The anisotropy of the magnetization is best examined through isothermal magnetization data \MH; such data for NQ crystals are shown in Fig.\,\ref{M_Compare}(b,c) (left panels) while data for Q-C crystals are shown in Fig.\,\ref{M_Compare}(e,f) (right panels).  The data suggest that the moments in \FGTx are easily polarized with relatively little anisotropy observed; easy-axis anisotropy with moments preferring to orient along [001] exists below at least 100\,K. This behavior appears to be independent of thermal history.  The effect of a demagnetizing field has been estimated for $H \parallel c$ using $N$=0.9 to obtain the internal field $H_{\textrm{int}}$ = $H_{\textrm{app}}$ -  $4 \pi N M_v$ where $M_v$ is the magnetization per unit volume and the multiplier $4 \pi$ is necessary for cgs units.\cite{DemagRectPlate}  Without applying a demagnetization correction, \FGTx appears to have essentially zero anisotropy.   Even with the correction, the anisotropy is somewhat small and increases when all of the Fe sublattices are ordered.  The easy-axis magnetism of Fe$_{4.7}$GeTe$_2$ crystals has an anisotropy field on the order of 1\,T at 2\,K. It is reasonable to speculate that this may change significantly between samples if the Fe-content varies.

In Fe$_{3-x}$GeTe$_2$, the anisotropy changes strongly with Fe content.\cite{May2016}  For Fe$_{3-x}$GeTe$_2$ compositions with small $x$, the anisotropy reaches 5\,T with easy-axis magnetism observed.\cite{Deiseroth2006,Chen2013}  While for large vacancy concentrations (Fe$_{2.75}$GeTe$_2$), the anisotropy is reduced to approximately 1\,T.  One goal for chemical manipulation of \FGTx would be to enhance the magnetic anisotropy, particularly at high $T$.  It would be interesting to see if this property could be tuned by extrinsic substitutions, as well as by controlling the Fe content.  

\subsection{Transport Properties}

\begin{figure}[ht!]%
\includegraphics[width=\columnwidth]{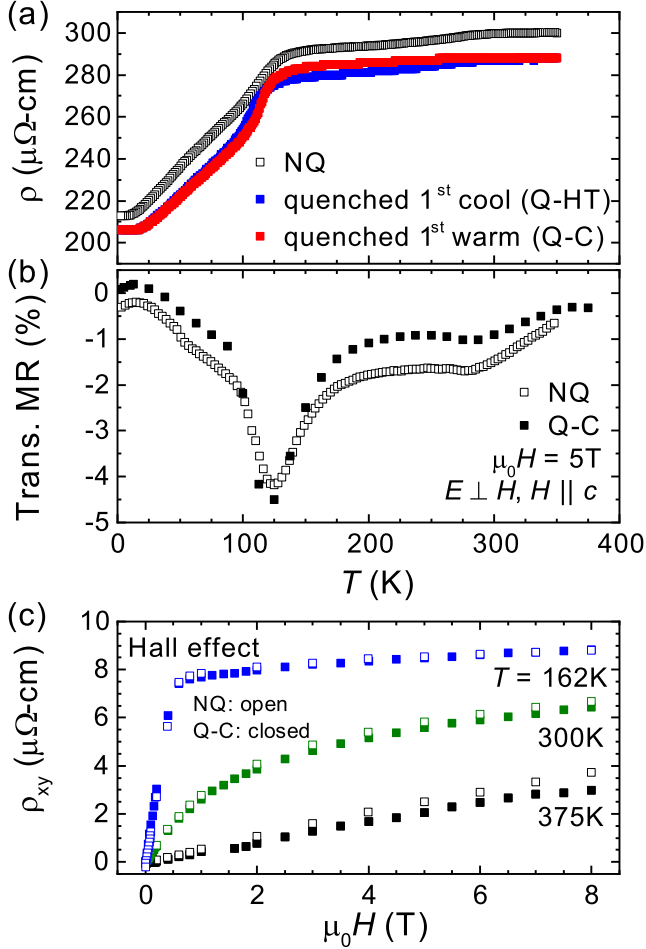}%
\caption{(color online). Comparison of the (a) in-plane electrical resistivity, (b) transverse magnetoresistance and (c) Hall effect in non-quenched (open symbols) and quenched-cycled (closed symbols) \FGTx\, single crystals.  Unlike the magnetization, little to no hysteresis is observed in the transport data and similar results are obtained regardless of the thermal history.}%
\label{Transport_Compare}%
\end{figure}

The transport properties of non-quenched and quenched single crystals are quite similar, as shown in Fig.\,\ref{Transport_Compare}.  All reported data are for transport within the basal plane ($\rho = \rho_{ab}$).  The electrical resistivity did not possess any significant hysteresis around the first-order transition between the Q-HT and Q-C phases, making transport appear rather different than the magnetization.  It would be interesting to see if a different behavior were observed for the resistance along the $c$ axis, since the structural transition clearly disrupts the layer stacking but not necessarily the in-plane structures.  The lack of hysteresis for in-plane transport is likely related to the lack of temperature dependence in $\rho$ above $\approx$ 120\,K.   This temperature dependence suggests that magnetic fluctuations on the Fe(1) sublattice provide the dominant temperature-dependent source of carrier scattering.   Such Fe(1) moment fluctuations seemingly exist in all \FGTx phases above 100-120\,K.  The magnetotransport data reported in this section further support this picture. The relatively large resistivity is likely further caused by static structural disorder that adds a temperature-independent scattering term that reduces the carrier lifetime.

As shown in Fig.\,\ref{Transport_Compare}(b), the magnetoresistance (MR) has a maximum at $\approx$120\,K, near where Fe(1) orders, and similar behavior is observed for the different \FGTx phases/crystals (NQ, Q-C).  The Hall effect data are also similar for the different types of crystals, as shown in Fig.\,\ref{Transport_Compare}(c).   The data in Fig.\,\ref{Transport_Compare}(b) are for transverse MR, with $H \parallel c$ and electric current flowing within the basal plane.   The magnetoresistance is determined by MR = $(\rho(H) - \rho(H=0))/\rho(H=0)$, and only the even contribution is included, $\rho(H)=(\rho(H) + \rho(-H))/2$.  The data in Fig.\,\ref{Transport_Compare}(b) for the NQ crystal are an exception because these MR results only utilized the value of $\mu_0H=+5$\,T to evaluate the temperature dependence of MR.   For all Hall effect data, only the odd contribution is included, $\rho_{xy}=(\rho_{xy}(H) - \rho_{xy}(-H))/2$.

Hall effect data are shown in Fig.\,\ref{Transport_Compare}(c) for both NQ and Q-C crystals. Quantitative agreement of the data is observed for these two crystals.  This further reveals that the thermal histories have little impact on the transport properties, though we are primarily concerned with the qualitative behaviors observed.  The same Q-C crystal was used to collect the Hall effect data in Fig.\,\ref{Transport_Compare}(c) and the MR data in Fig.\,\ref{ContourMaster_All}, as well as the Hall effect data in Figs.\,\ref{AHE_Simple}, \ref{AHE_Full}, \ref{sigma_AH}.  A six wire configuration was employed to collect MR and Hall effect data simultaneously, thus allowing a calculation of the Hall conductivity $\sigma_{xy}$.  Additional samples were also measured for comparison and the results were consistent with those shown.

\begin{figure}[ht!]%
\includegraphics[width=\columnwidth]{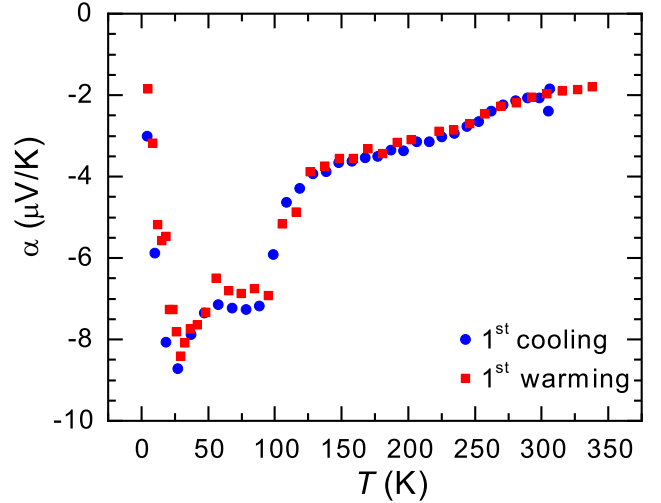}%
\caption{(color online) The Seebeck coefficient of a quenched \FGTx crystal as a function of temperature, with no notable hysteresis observed between first cooling (Q-HT) and first warming (Q-C).  The temperature gradient and voltage was within the basal plane.}%
\label{Seebeck}%
\end{figure}

The Hall effect data at and above room temperature suggest hole-like conduction (Hall coefficient $R_H > 0$).  However, as shown in Fig.\,\ref{Seebeck}, the Seebeck coefficient $\alpha$ is negative.  Like the Hall coefficient, the sign of the Seebeck coefficient is typically taken as an indication of the type of charge carrier ($\alpha < 0$ for $n$-type, $\alpha > 0$ for $p$-type in a single carrier system).  The behavior of $R_H$ and $\alpha$ thus suggest that both holes and electrons contribute to conduction.  Indeed, the temperature dependence of $\alpha$ is not consistent with a single carrier-type metal.  In a simple metal or heavily-doped semiconductor, the magnitude of $\alpha$ increases linearly with $T$.  In \FGTx, $\alpha$ is trending towards zero above $\approx$125\,K and this suggests a compensated system (contributions from holes and electrons cancel one another).   Upon cooling below 120\,K, the negative value of the Seebeck coefficient is increased and some sharp features are observed at lower $T$. $\alpha$ necessarily goes to zero at $T=0$.   These results suggest a change in the electronic structure occurs when the Fe(1) moments order.  While the Seebeck coefficient is senstivie to changes in carrier scattering, it is the energy dependence of the relaxation time that impacts the Seebeck coefficient and not the absolute value of the carrier's relaxation time.\cite{Fistul,Ziman63}  Thus, this effect is likely different from that driving the temperature depedence of the resistivity, even though they are both linked to magnetism on the Fe(1) sublattice.   A change in electronic structure is reasonable given that a magnetoelastic effect was previously observed around 100\,K via neutron powder diffraction.\cite{ACSNano}  It was demonstrated that upon cooling through 100\,K, the $a$ lattice parameter increases and the $c$ lattice parameter decreases.  The ratio $c/a$ increases upon cooling from 300\,K to $\approx$125\,K, decreases at the transition and then decreases upon cooling below 100\,K.  In light of the Seebeck coefficient results, there appears to be a strong coupling between the lattice, magnetism and electronic structure in \FGTx.  To complement these results, the temperature evolution of the magnetotransport is considered in detail below.    

\subsubsection{Magnetoresistance}

\begin{figure*}[ht!]%
\includegraphics[width=2\columnwidth]{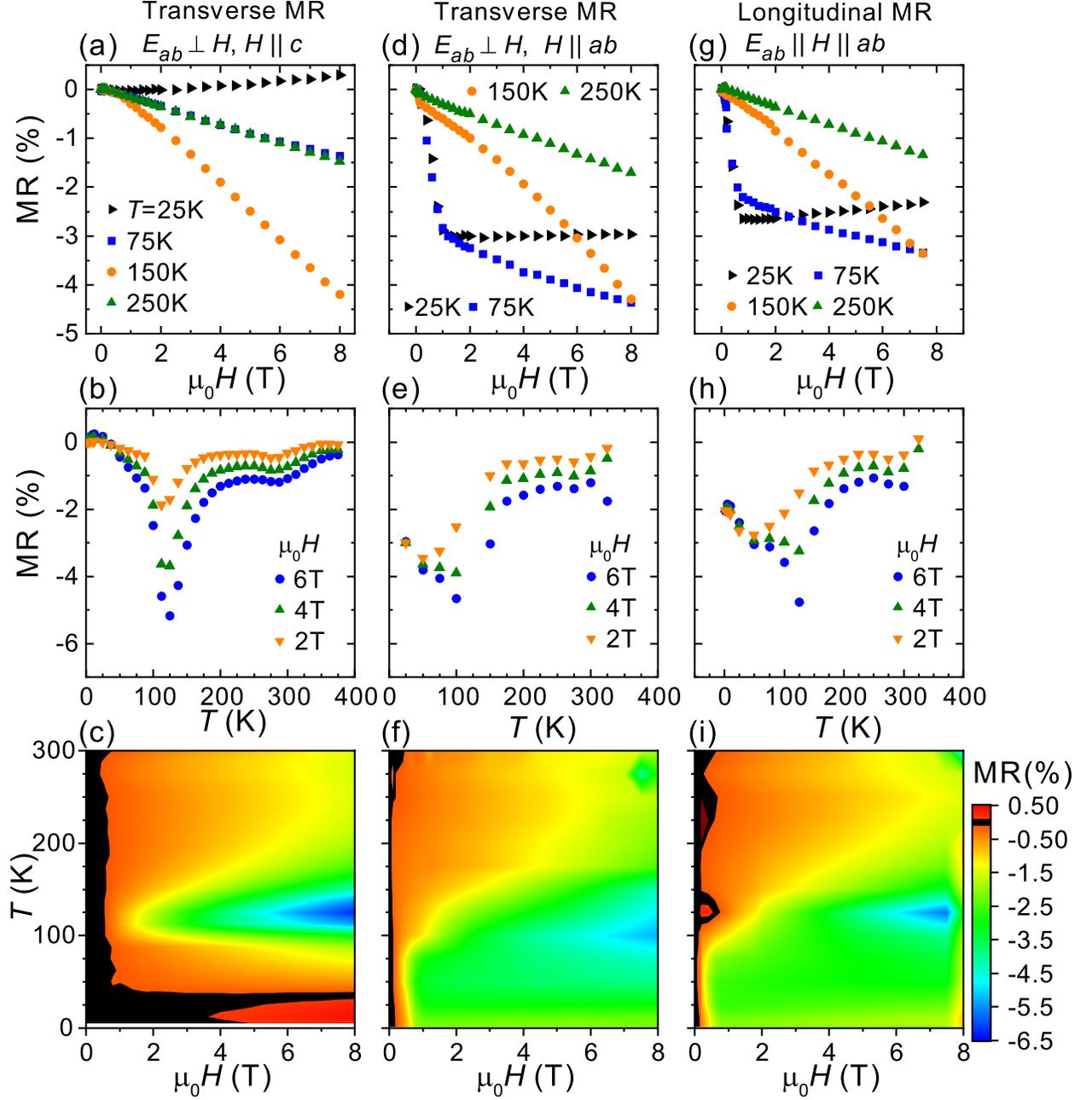}%
\caption{(color online). Comparison of magnetoresistance as percent change for two transverse orientations (left and center columns) and the longitudinal orientation (right column) with orientations of applied fields indicated at the top of each column.  In all cases, the current is flowing in the plane ($E_{ab}$).  For transverse MR, the magnetic field is either along the $c$-axis (a,b,c) or within the basal plane (d,e,f) but is necessarily perpendicular to the current.  For longitudinal MR, the current and magnetic field are parallel (g,h,i). Data are for a Q-C Fe$_{4.7}$GeTe$_2$ single crystal.}%
\label{ContourMaster_All}%
\end{figure*}

Various MR data are shown in Fig.\,\ref{ContourMaster_All}; data are for a Q-C crystal but similar results are expected for all crystal types based on the results in Fig.\,\ref{Transport_Compare}.  The three columns contain data for different orientations of the applied magnetic field.  In all cases, the electric field is within the plane $E_{ab}$.  For transverse MR, the magnetic field and electric current are perpendicular to one another.  Two transverse MR configurations were examined, defined by $H \parallel c$ (Fig.\,\ref{ContourMaster_All}(a,b,c)) and $H \parallel ab$ (Fig.\,\ref{ContourMaster_All}(d,e,f)).   The longitudinal MR has both the magnetic field and the electric current within the $ab$ plane and parallel to one another (Fig.\,\ref{ContourMaster_All}(g,h,i)).  For these three columns, the top row (panels a,d,g) presents the field-dependent data at a few temperatures. The middle row presents the temperature dependence at a few fixed fields.  The color of the contour plots in Fig.\,\ref{ContourMaster_All}(c,f,i) relates to the MR as indicated in the legend, with black indicating that the MR$\approx$0 ($|$MR$|<$0.1\%).  Data were collected down to 25\,K using steps of 25\,K, followed by measurements at 10, 5, and 2\,K.  The magnitude of the magnetic field was decreased from 8\,T by 0.5\,T down to 2\,T, then by 0.2\,T down to 0.2\,T, followed by a step size of 0.025\,T to $H$=0.

For all MR orientations, MR is predominantly negative and reaches a maximum magnitude near 120\,K.  A small peak in MR is observed near 300\,K (Fig.\,\ref{ContourMaster_All}(b)), suggesting that the loss of fluctuations at \TC have some impact on the MR.  Typically, MR will peak near the Curie temperature in a ferromagnet due to scattering by such fluctuations.  The maximum in MR near 120\,K is thus evidence of  the strong scattering of electrons by critical fluctuations on the Fe(1) magnetic sublattice.  For transverse MR with $H \parallel c$, the MR has a small and positive value at the lowest temperature and largest field.  This is likely due to loss of fluctuating moments at low $T$ (for $H=0$) combined with the normal increase in resistance due to Lorentz effects.

Saturation of the MR with increasing field is easily observed when $H \parallel ab$, for $T < 50\,K$, as shown in  Fig.\,\ref{ContourMaster_All}(d,g).  While MR never becomes positive for  $H \parallel ab$, the high-field slope of MR($H$) does become positive for low temperatures.  This indicates that a positive MR contribution (presumably from Lorentz forces) exists on top of the step-like drop in resistance that occurs when the moments are polarized into the basal plane by the applied field.  The step-like drop in resistance when $H \parallel ab$ is likely related to a reconfiguration of the Fermi surface upon reorienting the moments in this itinerant system.  The ordering of the Fe(1) sublattice clearly impacts the Fermi surface, as demonstrated through Hall and Seebeck coefficient data.  It is reasonable that the orientation of the moments may also impact the electronic structure.  Anisotropic electron-spin scattering may also be important to some degree.

\subsubsection{Hall Effect}

Upon cooling towards and below \TC, non-linearity in $\rho_{xy}(H)$ develops due to an anomalous Hall contribution (see Fig.\,\ref{Transport_Compare}(c)).  Multi-carrier conduction can also cause non-linearity in the field dependence of the Hall coefficient, though such field dependence is dictated by properties of the carriers.  For \FGTx, the clear evolution of an anomalous contribution with temperature suggests that magnetic polarization effects are influencing the Hall coefficient even above \TC.

\begin{figure}[ht!]%
\includegraphics[width=\columnwidth]{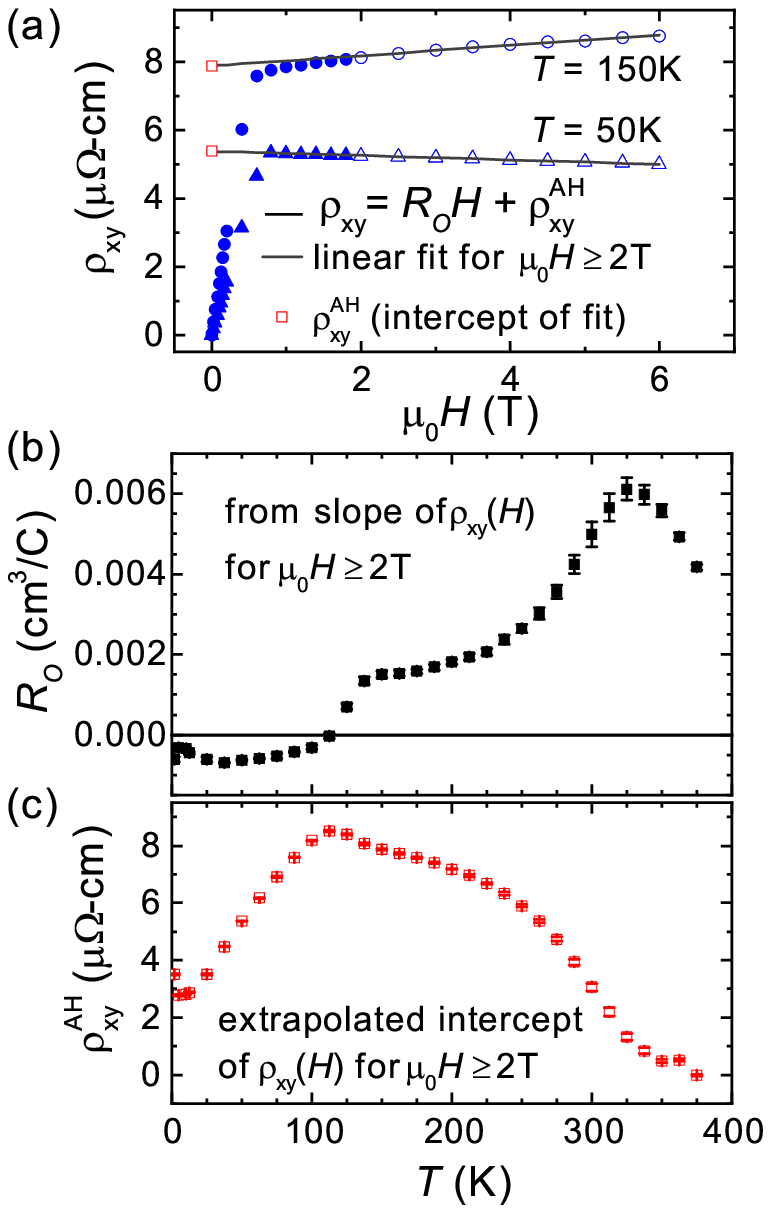}%
\caption{(color online). (a) Common method for analyzing Hall effect data demonstrating an anomalous contribution. Hall resistivity $\rho_{xy}$ data are shown for two characteristic temperatures, with a linear fit at high fields extrapolated to $H$=0 to demonstrate the anomalous contribution $\rho_{xy}^{AH}$.  (b) The ordinary Hall coefficient $R_O$  obtained from the slope of linear fits for $\mu_0H >$2\,T. (c) The anomalous Hall resistivity.  Error bars for (b,c) are from the non-weighted linear fits and are often smaller than the data marker.}%
\label{AHE_Simple}%
\end{figure}

When an anomalous Hall effect is present, the Hall coefficient is typically considered as a sum of an ordinary ($R_O$) and an anomalous contribution ($\rho^{AH}_{xy}$ or $R_S$).\cite{Nagaosa2010}  The Hall resistivity $\rho_{xy}$ is measured experimentally and a common representation is 

\begin{equation}
\rho_{xy} = R_O H + \rho^{AH}_{xy}.
\label{1}
\end{equation} 

\noindent The anomalous resistivity $\rho^{AH}_{xy}$ is typically obtained as the $H=0$ intercept of a linear fit to $\rho_{xy}$ versus $H$ at fields where $\rho_{xy}$ is linear in $H$ (when the induced magnetization is saturated). The slope of this linear fit provides the ordinary Hall coefficient, allowing access to the carrier density in the traditional manner when a single band model is appropriate. Such an analysis is demonstrated in Fig.\,\ref{AHE_Simple}(a). The anomalous Hall resistivity $\rho^{AH}_{xy}$ can be written as originating from an anomalous Hall coefficient ($R_S$) and the magnetization ($M$) parallel to the applied field, $\rho^{AH}_{xy} = R_S M$.  This allows one to address the impact of the magnetization on the Hall effect data, and one approach for such an analysis is to transform Eqn.\,\ref{1} into

\begin{equation}
\frac{\rho_{xy}}{H} = R_O + R_S\frac{M}{H}.
\label{2}
\end{equation} 

\noindent By plotting $\rho_{xy}/H$ versus $M/H$, the ordinary Hall coefficient is obtained from the intercept and the anomalous Hall coefficient from the slope.  This method can be more rigorous than utilizing Eqn.\,\ref{1} to obtain $R_O$, because it includes the field dependence of $M$ instead of assuming it is a constant.  $R_S$ can be positive or negative, regardless of $R_O$.  For \FGTx, both approaches give the same qualitative results; some quantitative differences in the ordinary Hall coefficient are observed.  

\begin{figure}[ht!]%
\includegraphics[width=\columnwidth]{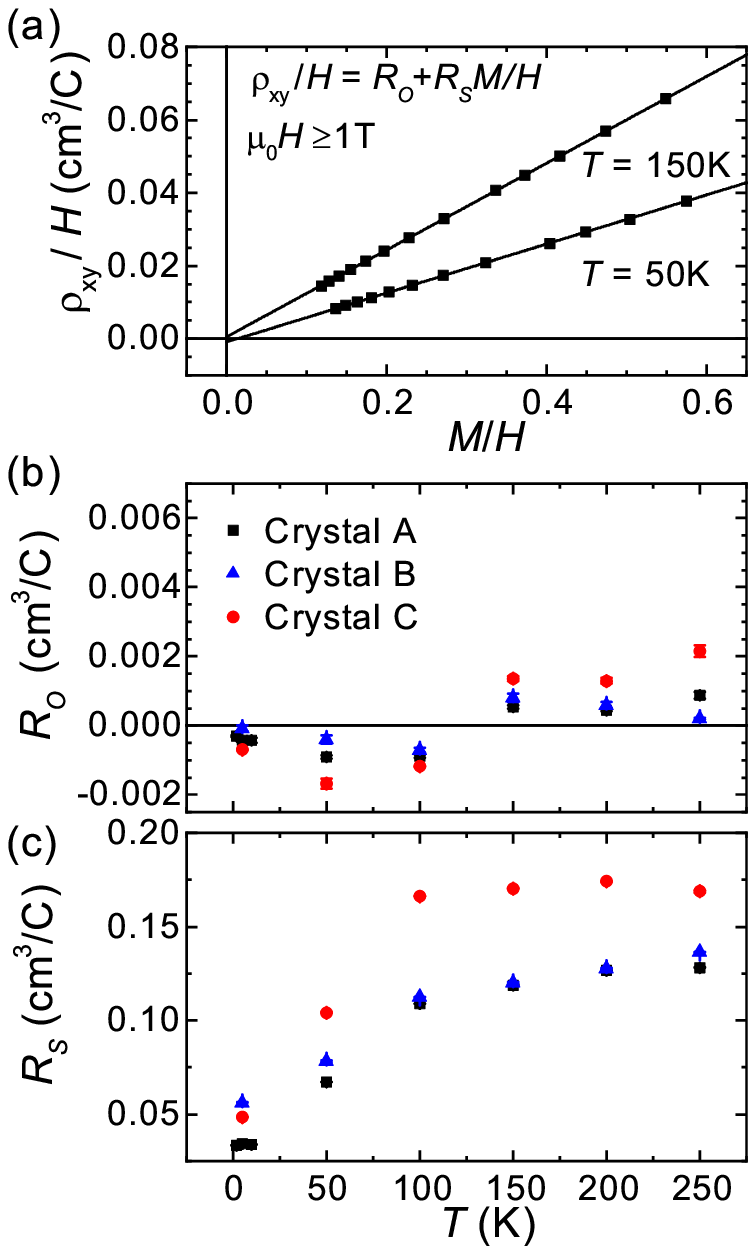}%
\caption{(color online). (a) Analysis of anomalous Hall effect data on Q-C crystals taking into account the impact of the $T$ and $H$ dependence of the magnetization $M$.  (b) The ordinary Hall coefficient $R_O$ from the high field limit $M/H$=0 (intercept) of the linear fits.  (c) The Anomalous Hall coefficient $R_S$ obtained from slopes of the linear fits.  Results for three samples are shown in (b,c), where error bars are from the non-weighted linear fits and are typically smaller than the data marker.}%
\label{AHE_Full}%
\end{figure}

The $\rho_{xy}$ data are first analyzed using the approach of Eqn.\ref{1}, as illustrated in Fig.\,\ref{AHE_Simple}(a).  As shown in Fig.\,\ref{AHE_Simple}(b), the ordinary Hall coefficient changes sign near 120\,K, indicating a dominance of electron-like contributions for $T <$ 120\,K.  Around this same temperature, the anomalous Hall resistivity reaches a maximum (Fig.\,\ref{AHE_Simple}(c)), as does the MR.   While the magnitude of the ordinary Hall coefficient obtained in this manner may be slightly impacted by the magnetization, the qualitative result is supported by the more detailed analysis that considers the impact of the induced magnetization (Eqn.\,\ref{2}).  The largest artifacts are likely observed near \TC, where strong polarization effects are observed in \MH without clear saturation.

The $\rho_{xy}$ data are analyzed using the approach of Eqn.\ref{2} in Fig.\,\ref{AHE_Full}.  The results are generally consistent with those reported in Fig.\,\ref{AHE_Simple}, and the change in sign of $R_O$ near 100\,K is again observed.  To illustrate reliability of these trends, results from measurements on three separate Q-C crystals are shown in Fig.\,\ref{AHE_Full}(b,c), where one set of \MH values were utilized but $\rho_{xy}$ data were collected for each crystal. Crystal A is the same as that utilized to generate the data shown in Figs.\,\ref{ContourMaster_All} and \ref{AHE_Simple}.  

The analysis method employing Eqn.\ref{2} is illustrated in Fig.\,\ref{AHE_Full}(a).  This approach yields an anomalous Hall coefficient, which is the proportionality between the anomalous resistivity and the magnetization.  When anomalous scattering effects are not important, this coefficient should be related to the band structure of the ordered state, and it is thus reasonable that $R_S$ is relatively independent of temperature for 100\,K $< T <$ 250\,K where ordered moments are well established. We note that the resistivity is also independent of $T$ in this region, and $R_S$ is typically proportional to $\rho_{xx}$ in a manner that depends on the model being considered.\cite{Nagaosa2010}   Below $\approx$100\,K, the Fe(1) sublattice orders and this causes a clear change in the band structure as indicated by the change in the sign of the ordinary Hall coefficient.  Upon cooling below $\approx$100\,K, $R_S$ begins to decrease, and this leads to the maximum in $\rho^{AH}_{xy}$ observed in Fig.\,\ref{AHE_Simple}(c).  We note that the same results are obtained when $H$ is replaced with $B$ in Eqn.\ref{2} and the effects of the internal and demagnetizing fields are incorporated into the analysis.

From a theoretical perspective, the anomalous Hall conductivity $\sigma^{AH}_{xy}$ is  typically considered.  When intrinsic, this conductivity is related to Berry curvature in the electronic structure and can be addressed from first principles calculations.  It is thus a property of topological significance.  Experimentally, it is difficult to determine when $\sigma^{AH}_{xy}$ is due to intrinsic electronic structure effects as opposed to scattering effects.\cite{Nagaosa2010}  Due to the nature of the conductivity matrix, the longitudinal resistivity $\rho_{xx}$ impacts the Hall conductivity:  $\sigma^{AH}_{xy}$ = $\rho^{AH}_{xy}$ /(($\rho_{xx}$)$^2$+($\rho^{AH}_{xy}$)$^2$) (and this is usually $\approx$ $\rho^{AH}_{xy}$ /($\rho_{xx}$)$^2$).  Since the anomalous Hall conductivity is nominally an $H$=0 effect in a ferromagnet, we utilize the $H$=0 values for $\rho_{xx}$ and $\rho^{AH}_{xy}$ to obtain $\sigma^{AH}_{xy}$.

\begin{figure}[ht!]%
\includegraphics[width=\columnwidth]{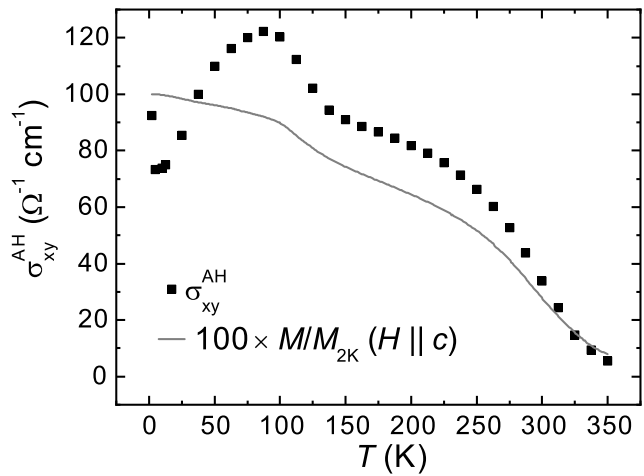}%
\caption{(color online) The anomalous Hall conductivity as a function of temperature (solid markers).  The solid curve plots the  temperature dependence of the magnetic moment measured along the $c$-axis for an applied field of 10\,kOe. The $M$ data are normalized to the value obtained at 2\,K ($\approx$2\,$\mu_B$/Fe).}%
\label{sigma_AH}%
\end{figure}

The anomalous Hall conductivity of Fe$_{4.7(2)}$GeTe$_2$ is plotted as a function of temperature in Fig.\,\ref{sigma_AH}.  $\sigma^{AH}_{xy}$ reaches a maximum just below the ordering temperature of the Fe(1) sublattice.  Above approximately 150\,K, the results reveal an order-parameter like increase that can be considered as driven by the magnetization.  It is difficult to know if the behavior at lower $T$ is intrinsic and driven by changes in the electronic structure that were evidenced by $R_O$ and $\alpha$ and are perhaps associated with  changes in the magnetization itself.  At this point, scattering effects and/or changes in $\rho_{xx}$ cannot the excluded.  We note that $\rho_{xy}^{AH}$ is not linear with either $\rho_{xx}$ or $\rho_{xx}^2$ below 125\,K where some temperature dependence of $\rho_{xx}$ is observed.  Thus, we can only say that the conductivity shown in Fig.\,\ref{sigma_AH} generally represents the maximum intrinsic anomalous Hall conductivity for Fe$_{4.7}$GeTe$_2$ crystals (within error).

The transport properties share some similarities with those observed in Fe$_{3-x}$GeTe$_2$.  In both of these ternary Fe-Ge-Te phases, a positive Hall coefficient and negative Seebeck coefficient are observed above \TC, demonstrating that both electrons and holes contribute to conduction.   The anomalous Hall contribution is positive for both materials.  The resistivity values at 300\,K are similar in both compounds and the scattering of carriers by vacancies is expected to be significant.  Magnetic fluctuations appear to provide the dominant temperature-dependent source for carrier scattering in both materials.  For flux-grown Fe$_{3-x}$GeTe$_2$ with $x \approx 0.25$, there is little temperature dependence of $\rho$ above \TC $\approx$ 150\,K.\cite{May2016} 

\section{Summary}

The thermal stability of quenched Fe$_{4.7(2)}$GeTe$_2$ crystals was examined by \textit{in-situ} diffraction and magnetization measurements to 700\,K as well as annealing studies, and a transition was observed at $T_{HT}\approx$550\,K. This is the critical temperature for a transition between a low-$T$ and high-$T$ crystal structure, though the microscopic change in structure remains unclear.  Samples quenched from above $T_{HT}$ are thus metastable, and upon cooling to cryogenic temperatures these crystals undergo a first-order transition below $\approx$100\,K to a state with higher \TC.  Based upon magnetization measurements, this state with enhanced \TC appears stable to at most $T_{HT}$.  The extent to which this instability remains in exfoliated samples would be interesting to determine since the transition at $T_{HT}$ appears to mainly affect layer stacking.   

The transport data provide a complementary perspective on the magnetic complexity of Fe$_{4.7(2)}$GeTe$_2$.  The in-plane transport data do not appear to be strongly impacted by thermal history.  This is consistent with the diffraction-based evidence that the in-plane crystal structure is not strongly impacted across $T_{HT}$.  The in-plane transport data reveal that dynamic moments on the Fe(1) sublattice strongly scatter charge carriers (above $\approx$120\,K).  A maximum in the magnetoresistance and anomalous Hall resistivity is observed near 120\,K, above which the resistivity has very little temperature dependence. The Hall and Seebeck coefficients reveal that both holes and electrons contribute to charge transport when Fe(1) moments are fluctuating.  However, it appears that hole conduction is eliminated or strongly suppressed when the Fe(1) moments order.  This reconstruction of the Fermi surface could be interesting to inspect using various spectroscopic techniques.  In this regard, it is worth noting that the magnetic ordering of the Fe(1) sublattice is coupled to the lattice via a magnetoelastic effect that is expected to be present, and reversible, in all Fe$_{4.7(2)}$GeTe$_2$ specimens.  In addition to employing other characterization techniques and theoretical calculations, it would be valuable to find a way to vary the Fe-content in crystals and explore how the physical properties and metastability depend on composition.  

\section{Acknowledgments}

This work was supported by the U. S. Department of Energy, Office of Science, Basic Energy Sciences, Materials Sciences and Engineering Division. High-temperature x-ray diffraction experiments (C.A.B.) were sponsored by the Laboratory Directed Research and Development Program of Oak Ridge National Laboratory, managed by UT-Battelle, LLC, for the U. S. Department of Energy.   We thank Brian Sales and Satoshi Okamoto for useful discussions.


%

\end{document}